\documentstyle[psfig]{l-aa}

\begin{document}
\hyphenation{Hobbs Maz-zi-tel-li PN SN}

   \thesaurus{10         
              (10.01.1;  
	       10.05.1;  
               10.19.1)} 
   \title{The stellar origin of $^7$Li}

   \subtitle{Do AGB stars contribute a substantial fraction of the 
	     local Galactic lithium abundance?}

   \author{Donatella Romano
           \inst{1},
	   Francesca Matteucci
	   \inst{2, 1},
	   Paolo Ventura
	   \inst{3}
	   \and
	   Francesca D'Antona
	   \inst{3}
          }

   \offprints{D. Romano}

   \institute{\inst{1} International School for Advanced Studies, SISSA/ISAS, 
	      Via Beirut 2--4, 34014 Trieste, Italy\\
	      \inst{2} Dipartimento di Astronomia, Universit\`a di Trieste,
	      Via G.B. Tiepolo 11, 34131 Trieste, Italy\\
	      \inst{3} Osservatorio Astronomico di Roma, 
	      Via Frascati 33, 00040 Monte Porzio Catone, Italy
             }

   \date{Received 22 December 2000/ Accepted 28 May 2001}

   \maketitle
   \markboth{D. Romano et al.\,: The stellar origin of $^7$Li}{}

   \begin{abstract}

	We adopt up-to-date $^7$Li yields from asymptotic giant branch stars 
	in order to study the temporal evolution of $^7$Li in the solar 
	neighbourhood in the context of a revised version of the two-infall 
	model for the chemical evolution of our galaxy. 

	We consider several lithium stellar sources besides the asymptotic 
	giant branch stars such as Type II supernovae, novae, low-mass giants 
	as well as Galactic cosmic rays and low-mass X-ray binaries.

	We conclude that asymptotic giant branch stars cannot be considered 
	as important $^7$Li producers as believed in so far and that the 
	contribution of low-mass giants and novae is necessary to reproduce 
	the steep rise of the $^7$Li abundance in disk stars as well as the 
	meteoritic $^7$Li abundance. Lithium production in low-mass X-ray 
	binaries hardly affects the temporal evolution of $^{\mathbf 7}$Li in 
	the solar neighbourhood.

   \end{abstract}

\section{Introduction}

   The upper envelope of the observed log\,$\epsilon$($^7$Li) vs. 
   [Fe/H]\footnote{Here we use the notation [X/H] = 
   log\,(X/H)$_{\mathrm{star}}$ $-$ log\,(X/H)$_\odot$ and log\,$\epsilon$(X) 
   = log\,(N$_{\mathrm{X}}$/N$_{\mathrm{H}}$) + 12, where X and H are the 
   abundances by mass and N$_{\mathrm{X}}$ and N$_{\mathrm{H}}$ are the number 
   of atoms of the element X and hydrogen, respectively.} 
   diagram is generally believed to reflect the $^7$Li enrichment history of 
   the interstellar medium (ISM) in the solar neighbourhood. Therefore, it can 
   be used to constrain models of Galactic chemical evolution aiming at 
   explaining the temporal evolution of this element. 
   The major features of the observational diagram are {\it i)} a large 
   plateau at low metallicities ({\it lithium-metallicity plateau}, firstly 
   pointed out by Rebolo, Molaro, \& Beckman 1988) followed by {\it ii)} a 
   steep rise afterwards. 

   A large spread is observed in the data for Population I dwarfs, increasing 
   with metallicity and commonly interpreted as a signature of processes of 
   lithium dilution and/or destruction in stars. 

   Field halo dwarfs hotter than 5700 K exhibit a weaker dependence on 
   effective temperature, $T_{\mathrm{eff}}$, and metallicity, [Fe/H], and a 
   smaller dispersion than seen in Population I stars. 
   However, ultra-Li-deficient, warm, halo stars do exist (Spite, Maillard, 
   \& Spite 1984; Hobbs \& Mathieu 1991; Hobbs, Welty, \& Thorburn 1991; 
   Thorburn 1992; Thorburn \& Beers 1993; Spite et al. 1993; Ryan et al. 
   2001a), complicating the overall picture. Actually, there is vigorous 
   debate about the existence and magnitude of any dispersion, and there are 
   active controversies about trends with $T_{\mathrm{eff}}$ and [Fe/H] 
   (Thorburn 1994; Molaro, Primas, \& Bonifacio 1995; Ryan et al. 1996; Spite 
   et al. 1996; Bonifacio \& Molaro 1997). In particular, the existence of 
   trends of $^7$Li with metallicity would be a signature of Galactic chemical 
   evolution.

   By comparing theoretical stellar evolution models with the observational 
   Population II lithium data one can obtain bounds on stellar lithium 
   depletion in halo stars. This has implications for stellar structure, 
   Galactic chemical evolution, and Big Bang nucleosynthesis (BBN). 
   From a combination of the dispersion in the data, the detection of the 
   fragile isotope $^6$Li in some halo stars (Smith, Lambert, \& Nissen 1993, 
   1998; Hobbs \& Thorburn 1994, 1997; Cayrel et al. 1999; Nissen et al. 1999, 
   2000), and the flatness of the halo plateau a depletion at the 
   0.15\,--\,0.2 dex level, with a firm upper bound of 0.5\,--\,0.6 dex, has 
   been suggested (Pinsonneault, Charbonnel, \& Deliyannis 2000; Pinsonneault 
   et al. 1999). In this scenario, the primordial lithium abundance should be 
   (slightly) higher than what we effectively see. 

   However, one could ask whether there is any depletion at all: are we 
   directly seeing the primordial lithium abundance ({\it e.g.}, Bonifacio \& 
   Molaro 1997), or is the primordial lithium abundance lower than the 
   observed plateau value because of a significant early contribution from 
   Galactic chemical evolution (Ryan, Norris, \& Beers 1999; Ryan et al. 2000; 
   Suzuki, Yoshii, \& Beers 2000)? 
   It seems that consistency of the BBN predictions and light element 
   abundances hinges on some depletion of $^7$Li in old halo stars (Burles, 
   Nollett, \& Turner 2000), although in our view a clear answer is still 
   precluded. For this reason we will investigate the effect of adopting three 
   different values for the primordial lithium abundance in the chemical 
   evolution code.

   \begin{figure*}
\hfill    \psfig{figure=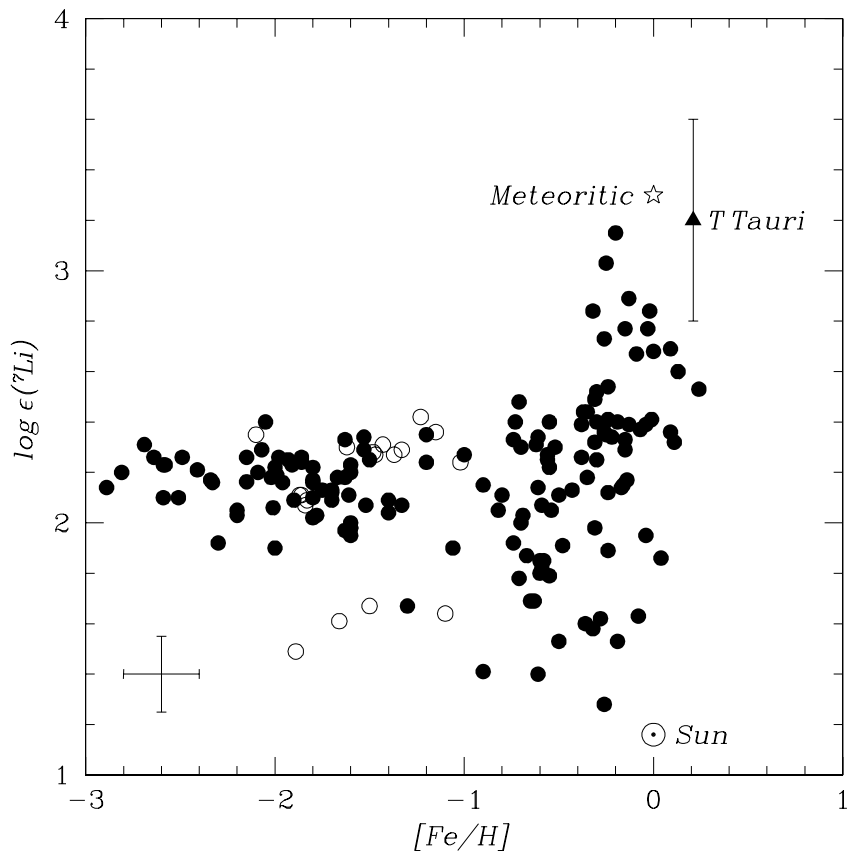,width=12cm}
\hfill    \parbox[b]{5.5cm}{\caption{The observational log\,$\epsilon$($^7$Li) 
				     vs. [Fe/H] diagram. {\it Filled circles}: 
				     data from the compilation of Romano et 
				     al. (1999); {\it open circles}: new 
				     measurements by Ryan et al. (2001b); both 
			             symbols are referring to objects with 
			             $T_{\mathrm{eff}}$ $\ge$ 5700 K. 
				     The solar abundance is taken from Anders 
				     \& Grevesse (1989), the meteoritic one is 
				     from Nichiporuk \& Moore (1974). 
				     The $^7$Li abundance in T\,Tauri stars we 
				     show is the one suggested by 
				     Stout-Batalha, Batalha, \& Basri (2000), 
				     at 2-$\sigma$ error.
                           }}%
      \label{FigDat}%
   \end{figure*}

   In the standard framework of the origin and evolution of the light 
   elements, $^7$Li is produced in significant abundance ($\sim$ 1/10 solar) 
   in BBN. The processes of spallation of Galactic cosmic rays (GCRs) composed 
   of $p$, $\alpha$ and C, N, and O on interstellar C, N, and O and $p$, 
   $\alpha$ nuclei, respectively, joint to an $\alpha$-$\alpha$ fusion 
   channel are responsible for $\sim$ 20\,--\,30\,\% of the local Galactic 
   abundance of $^7$Li ({\it e.g.}, Reeves, Fowler, \& Hoyle 1970; Meneguzzi, 
   Audouze, \& Reeves 1971). The remaining unexplained fraction of the solar 
   abundance has to originate from a stellar source. 

   Different studies trying to explain the observed log\,$\epsilon$($^7$Li) 
   vs. [Fe/H] trend by taking into account different astrophysical sites for 
   $^7$Li production have appeared in the literature ({\it e.g.}, D'Antona \& 
   Matteucci 1991; Matteucci, D'Antona, \& Timmes 1995; Abia, Isern, \& Canal 
   1995; Romano et al. 1999, hereafter Paper I; Casuso \& Beckman 2000; Ryan 
   et al. 2001b). However, most of them focused on two or three categories of 
   $^7$Li producers alone, while neglecting the others. The main purpose of 
   this work is to study the interplay between several categories of $^7$Li 
   producers in order to explain both the rise from the lithium-metallicity 
   plateau and the meteoritic abundance of $^7$Li; in particular, to test the 
   effect of new detailed nucleosynthesis calculations by Ventura, D'Antona \& 
   Mazzitelli(2000) on the $^7$Li produced in asymptotic giant branch (AGB) 
   stars. To do that, we adopt a chemical evolution model which already 
   reproduces the majority of the observational constraints in the Milky Way 
   (Chiappini, Matteucci, \& Gratton 1997; Chiappini, Matteucci, \& Romano 
   2001).

   In Section 2 we describe the chemical evolution model and the various 
   lithium producers. In Section 3 the results are presented and in Section 4 
   some conclusions are drawn.

\section{The model}

\subsection{Basic assumptions}

   The adopted model is a revised version of the two-infall model of Chiappini 
   et al. (1997), especially devised in order to study the radial properties 
   of the Galactic disk (Chiappini et al. 2001). 

   The Galaxy is assumed to form out of two main episodes of accretion: 
   during the first one the halo/thick-disk forms; during the second one 
   the thin-disk is built up, mainly out of matter of primordial chemical 
   composition, except for some fraction of enriched gas from the halo. 
   The stellar generations forming during the Galactic lifetime distribute 
   their stellar masses according to a Scalo initial mass function (IMF). The 
   IMF is kept constant in space and time. The star formation, proportional 
   to both the gas surface density and the total mass surface density at every 
   time, is a self-regulated process, in the sense that it stops when the gas 
   surface density falls below a certain critical threshold and it starts 
   again when an above-threshold value is recovered due to infall of 
   primordial matter coupled to restoration of gaseous matter by dying stars.

   As far as the solar vicinity is concerned, the major differences with 
   the Chiappini et al. (1997) model which we adopted in Paper I are the 
   following: the sun is located at a distance of 8 kpc from the Galactic 
   center, rather than 10 kpc; the adopted Galactic age is 14 Gyr, rather 
   than 15 Gyr; the timescales for mass accretion in the halo/thick-disk and 
   thin-disk components are 0.8 and 7 Gyr, respectively (rather than 2 and 8 
   Gyr). 
   We refer the reader to Chiappini et al. (1997, 2001) for a full description 
   of model parameters and results. Here we will only address the aspects more 
   strictly related to $^7$Li evolution (production in different astrophysical 
   environments and astration in stars).

\subsection{Lithium synthesis prescriptions}

   The primordial $^7$Li abundance is set to be log\,$\epsilon$($^7$Li)$_P$ = 
   2.2 dex (Bonifacio \& Molaro 1997). The cases log\,$\epsilon$($^7$Li)$_P$ = 
   2.09 dex (Ryan et al. 2000) and 2.4 dex ({\it e.g.}, Pinsonneault et al. 
   2000) are also investigated (see discussion in Section 1).

   Lithium astration in stars of all masses is taken into account in a simple 
   way, by assuming that each stellar generation fully destroyes the lithium 
   present in the progenitor gas out of which it forms.

   As far as the $^7$Li synthesis is concerned, we include in the chemical 
   evolution code the following $^7$Li sources: the $\nu$-process in Type II 
   supernovae (SNe), AGB stars which undergo the hot bottom burning (HBB) 
   process, low-mass red giants, novae, and GCRs. Moreover, we try to argue 
   whether low-mass X-ray binaries (LMXBs) could play a role regarding the 
   production of $^{\mathbf 7}$Li at a Galactic scale. The hypothesis of 
   instantaneous recycling is relaxed and the stellar lifetimes are taken into 
   account in great detail: according to this, different lithium sources 
   restore their newly synthesized $^7$Li on different timescales. This is an 
   important point to stress, since the shape of the growth in the 
   log\,$\epsilon$($^7$Li) vs. [Fe/H] diagram can be justified just in terms 
   of timescales of $^7$Li production.

\subsubsection{Type II supernovae}

   The neutrino flux emitted from a cooling proto-neutron star alters the 
   traditional outcome of explosive nucleosynthesis from Type II SNe. $^7$Li 
   production takes place mostly in the helium shell. The key reaction for 
   producing $^7$Li is the excitation of $^4$He by $\mu$- and $\tau$-neutrinos 
   (and their anti-neutrinos) through inelastic scattering, followed by 
   nuclear de-excitation via emission of a neutron or a proton. The decay 
   products then react with the abundant $^4$He to produce $^7$Li and $^7$Be 
   (which decays later to $^7$Li) (see Hartmann et al. 1999 and references 
   therein). 

   Unfortunately, no observational evidence exists supporting that Type II SNe 
   are able to enrich the ISM in $^7$Li. Therefore, the estimated theoretical 
   yields are very uncertain. In particular, they are mostly sensitive to the 
   properties of the neutrino flux. $\nu$-process yields for a model grid in 
   stellar mass and metallicity have been computed by Woosley \& Weaver 
   (1995). We included $^7$Li + $^7$Be yields from Woosley \& Weaver in a 
   model for the chemical evolution of the solar neighbourhood and showed 
   that, in order to reproduce the extension of the lithium-metallicity 
   plateau, we had to require a reduction of these yields by at least a 
   factor of 2 (Paper I). Our claim was in agreement with the upper limit on 
   the contribution to $^7$Li production by $\nu$-induced nucleosynthesis 
   suggested by Ramaty et al. (1997). Moreover, observations of Be and B in 
   stars covering 3 orders of magnitude in metallicity ($-$\,3.0 $\le$ [Fe/H] 
   $\le$ $+$\,0.0) suggest a scenario for light element production which 
   argues against the majority of B being produced in the $\nu$-process 
   (Duncan et al. 1997), thus putting another, independent constraint on the 
   $\nu$-process yields.

   In the light of the above considerations, we will take here as $^7$Li + 
   $^7$Be yields from Type II SNe those by Woosley \& Weaver (1995) reduced 
   to a half, although an even larger reduction factor could be required 
   (Vangioni-Flam et al. 1996).

   \begin{table}
      \caption{$^7$Li yields from AGB stars. The yield is the fraction of 
               the initial mass of the star which is ejected as newly produced 
               $^7$Li during the whole stellar lifetime.}
         \label{TabYields}
      \[
         \begin{tabular}{c c c c c}
         \multicolumn{3}{c}{}\\
            \hline
            \noalign{\smallskip}
	    $M_{init}(M_\odot)$  &  $^7$Li yield ($\eta_R$ = 0.01)  
			         &  $^7$Li yield ($\eta_R$ = 0.10)\\
            \noalign{\smallskip}
            \hline
            \noalign{\smallskip}
		$Z$ = 0.02 & & \\
            \noalign{\smallskip}
            \hline
            \noalign{\smallskip}
    3.5   &   1.28E-11          &	  5.07E-11  \\
    4.0   &   3.34E-10          &	  1.78E-09  \\
    4.5   &   6.25E-10          &	  6.53E-09  \\
    5.0   &   1.41E-09          &	  1.09E-08  \\
    5.5   &   3.11E-09          &	  1.87E-08  \\
    6.0   &   5.21E-09          &	  2.47E-08  \\
            \noalign{\smallskip}
            \hline
            \noalign{\smallskip}
		$Z$ = 0.01 & & \\
            \noalign{\smallskip}
            \hline
            \noalign{\smallskip}
    3.3   &   4.03E-10          &	     --     \\
    3.5   &   4.03E-10          &	  1.43E-11  \\
    4.0   &   6.08E-10          &	  4.94E-09  \\
    4.5   &   7.29E-10          &	  6.49E-09  \\
    5.0   &   1.29E-09          &	  9.61E-09  \\
    5.5   &   1.89E-09          &	  1.69E-08  \\
    6.0   &   4.10E-09          &	  2.31E-08  \\
            \noalign{\smallskip}
            \hline
            \noalign{\smallskip}
		$Z$ = 0.004 & & \\
            \noalign{\smallskip}
            \hline
            \noalign{\smallskip}
    2.2   &   1.71E-11          &       \\
    2.5   &   1.70E-11          &       \\
    3.0   &   5.00E-10          &       \\
    3.5   &   3.65E-10          &       \\
    4.0   &   3.60E-10          &       \\
    4.5   &   3.45E-10          &       \\
    5.0   &   4.17E-10          &       \\
    5.5   &   5.09E-10          &       \\
            \noalign{\smallskip}
            \hline
            \noalign{\smallskip}
		$Z$ = 0.001 & & \\
            \noalign{\smallskip}
            \hline
            \noalign{\smallskip}
    3.0   &   4.12E-10          &	  1.34E-11  \\
    3.5   &   2.58E-10          &	  2.21E-10  \\
    4.0   &   1.97E-10          &	  2.22E-09  \\
    4.5   &   1.20E-10          &	  2.14E-09  \\
    5.0   &   2.73E-10          &	  2.29E-09  \\
    5.5   &   2.47E-10          &	  2.07E-09  \\
            \noalign{\smallskip}
            \hline
            \noalign{\smallskip}
		$Z$ = 0.0006 & & \\
            \noalign{\smallskip}
            \hline
            \noalign{\smallskip}
    2.5   &   1.05E-11          &       \\
    3.0   &   3.91E-10          &       \\
    3.5   &   2.26E-10          &       \\
    4.0   &   1.92E-10          &       \\
    4.5   &   1.40E-10          &       \\
    5.0   &   1.20E-10          &       \\
    5.5   &   1.80E-10          &       \\
            \noalign{\smallskip}
            \hline
            \noalign{\smallskip}
		$Z$ = 0.0002 & & \\
            \noalign{\smallskip}
            \hline
            \noalign{\smallskip}
    2.5   &   2.01E-10          &       \\
    3.0   &   4.20E-10          &       \\
    3.5   &   2.69E-10          &       \\
    4.0   &   1.75E-10          &       \\
    4.5   &   1.36E-10          &       \\
    5.0   &   1.14E-10          &       \\
            \noalign{\smallskip}
            \hline
         \end{tabular}
      \]
   \end{table}

   \begin{figure}
			   \psfig{figure=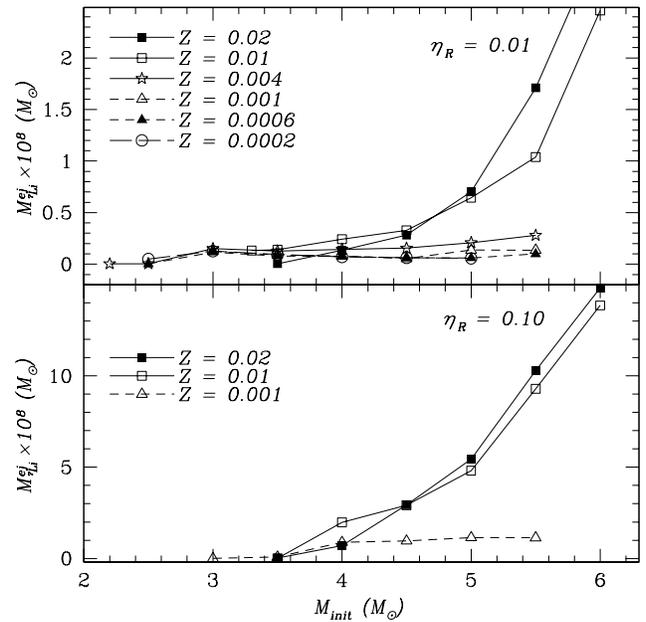,width=8.8cm}
			   \caption{Mass ejected in the form of newly produced 
                                    $^7$Li during the whole stellar lifetime 
                                    as a function of the initial mass of the 
                                    star, at various metallicities and for two 
                                    choices of the Reimer's parameter of mass 
                                    loss, $\eta_R$. 
                                    $^7$Li production occur along the AGB, 
                                    during the termally pulsing phase.}
      \label{FigYields}%
   \end{figure}

\subsubsection{AGB stars}

   AGB stars have been recognized to be $^7$Li producers since long ago (Smith 
   \& Lambert 1989, 1990; Sackmann \& Boothroyd 1992). According to Sackmann 
   \& Boothroyd, $^7$Li production would occur in the mass range $M$ $\sim$ 
   4\,--\,6 $M_\odot$, when the temperature at the base of the convective 
   envelope exceeds 50 $\times$ 10$^6$ K and the $^7$Be transport mechanism 
   (Cameron \& Fowler 1971) works. Under these conditions, $^7$Li abundances 
   as large as log\,$\epsilon$($^7$Li) $\sim$ 4\,--\,4.5 can be achieved in 
   the external layers, in principle allowing these stars to efficiently 
   pollute the ISM via mass loss and/or planetary nebula (PN) ejection. A 
   metallicity effect would also prevent these stars from contributing 
   significant amounts of newly synthesized $^7$Li at early times (Plez, 
   Smith, \& Lambert 1993; see also discussions in Matteucci et al. 1995 and 
   Paper I), thus helping to keep the lithium-metallicity plateau flat until 
   an ISM metallicity of [Fe/H] $\sim$ $-$\,0.5 dex is achieved, in agreement 
   with observations (Fig.~\ref{FigDat}). On these grounds, massive AGBs were 
   proposed as a significant source of lithium Galactic enrichment ({\it 
   e.g.}, D'Antona \& Matteucci 1991; Matteucci et al. 1995), even if a 
   reliable computation of the $^7$Li yields was still in demand at that time.

   Very recently such a computational task has been accomplished by means of 
   the code ATON 2.0 (Ventura et al. 1998). This computational tool is 
   especially devised to follow the AGB HBB phase: it includes diffusive 
   mixing coupled to nuclear evolution and it treats turbulence in the Full 
   Spectrum of Turbulence model (Canuto \& Mazzitelli 1991; Canuto, Goldman \& 
   Mazzitelli 1996). Results for $Z$ = 0.01 can be found in Ventura et al.
   (2000). Here we enlarge the model grid to a wide range of metallicities 
   ($Z$ = 0.0002, 0.0006, 0.001, 0.004, and 0.02) and include it in a code for 
   the chemical evolution of the solar neighbourhood. The complete, enlarged 
   model grid is shown in Table~\ref{TabYields} and Fig.~\ref{FigYields} for 
   two choices of the Reimer's factor $\eta_R$ entering Bl\"ocker's (1995) 
   prescription for mass loss. Observational evidence suggests that the lower 
   value of $\eta_R$ should be preferred, at least at low metallicities 
   (Ventura et al. 2000).

   Lithium production during the AGB phase is very efficient, but in the last 
   phases of evolution the fast $^3$He consumption implies a soon decrease in 
   the $^7$Li abundance. As a result, the net yields are too low to 
   significantly contribute to $^7$Li enrichment on a Galactic scale (see 
   Section 3 for detailed chemical evolution results).

\subsubsection{Low-mass red giants}

   Population I red giants usually have lithium abundances lying in the range 
   $-$\,1 $<$ log\,$\epsilon$($^7$Li) $<$ 1 (Lambert, Dominy, \& Sivertsen 
   1980; Brown et al. 1989). However, a few ($\sim$ 1\,\%) of them show 
   lithium abundances far in excess of standard predictions ({\it e.g.}, 
   Wallerstein \& Sneden 1982; Hanni 1984; Brown et al. 1989; da Silva, de la 
   Reza, \& Barbuy 1995), occasionally having abundances much higher than the 
   present ISM value ({\it e.g.}, de la Reza, Drake, \& da Silva 1996; 
   Balachandran et al. 2000). Frequently, these Li-rich red giants have large 
   infrared excesses, which are interpreted as associated circumstellar dust 
   shells. Interestingly enough, these Li-rich giants have not yet reached the 
   AGB, and at least some of them are observed to be low-mass stars; thus they 
   cannot have experienced $^7$Li creation via HBB, and do not enter in the 
   group we have analyzed in detail in the previous paragraphs.

   A scenario has been proposed in which all low-mass giants ($M$ $<$ 2.5 
   $M_\odot$) suffer a prompt $^7$Li enrichment in the upper part of the red 
   giant branch (RGB), after the first dredge-up and before the RGB tip. A 
   circumstellar dusty shell enriched with $^7$Li forms, which detaches later 
   from the star when the internal mechanism of $^7$Li production ceases. 
   In this way, the ISM is polluted by newly synthesized $^7$Li (de la Reza 
   et al. 1996, 1997). Using this stellar internal prompt $^7$Li enrichment 
   scenario coupled to mass loss, de la Reza et al. (2000) have explored the 
   possibility of $^7$Li enrichment by low-mass, metal-poor RGB stars in 
   globular clusters and suggested them as a potential source of $^7$Li 
   enrichment in the Galactic disk.

   Maybe the best mechanism for producing the $^7$Li photospheric enrichment 
   for these low-mass giants is the cool bottom process (CBP) based on $^7$Be 
   production from $^3$He in the H-burning shell, followed by transport of the 
   fresh $^7$Be up to the base of the convective layer to be then taken to the 
   cooler stellar surface where it decays to $^7$Li (Sackmann \& Boothroyd 
   1999). In this scenario, a discontinuous $^7$Li enrichment is linked to 
   mass loss, and the process can be repeated depending on the availability of 
   $^3$He. Contrary to AGB stars, where $^7$Be is quickly mixed away by 
   ordinary convection, in RGB stars some extra-mixing is required. 
   Denissenkov \& Weiss (2000) have proposed a mechanism which is able to 
   trigger the extra-mixing: a giant planet (or brown dwarf) is engulfed by a 
   red giant; this external event activates inside the giant the $^7$Be 
   transport mechanism which results in producing $^7$Li. The great advantage 
   of this solution is that it can account not only for the Li-production but 
   also for the subsequent Li-depletion, on a quick timescale consistent with 
   the results of de la Reza et al. (1996). The amount of $^7$Li produced in a 
   single episode of lithium enrichment can exceed log\,$\epsilon$($^7$Li) 
   $\sim$ 4, but it is critically related to the details of the extra-mixing 
   mechanism (Sackmann \& Boothroyd 1999). 

   The amount of $^7$Li injected into the ISM depends strongly on the 
   parameters of the lithium enrichment scenario: the size and timescales of 
   the lithium enhancements, the magnitude and timescales of the mass loss, 
   and whether, how often, and at what points on the RGB such enhancement 
   episodes occur. 
   de la Reza et al. (1996, 1997), with log\,$\epsilon$($^7$Li) $\le$ 4 for 
   timescales $\le$ 10$^5$ yr, mass loss in the range 
   10$^{-7}$\,--\,10$^{-10}$ $M_\odot$ yr$^{-1}$, and recurrence of $\sim$ 10 
   times per star at log\,$L$ $\sim$ 2, predict an average $^7$Li abundance in 
   the total amount of material ejected from low-mass stars less than $\sim$ 
   1\,\% of the cosmic abundance, a negligible quantity. However, if at least 
   some $^7$Li enrichment episodes occur near the tip of the RGB, this average 
   $^7$Li abundance could be more than an order of magnitude larger, yielding 
   to a non-negligible $^7$Li enrichment of the ISM (Sackmann \& Boothroyd 
   1999; but see also Charbonnel \& Balachandran 2000).

   Here we investigate this latter scenario, more favourable to $^7$Li 
   production. We assume that: {\it i)} all stars in the mass range $M$ = 
   1\,--\,2 $M_\odot$ experience enhanced RGB lithium abundances in 
   conjunction with episodic mass loss, for a period lasting $\sim$ 10$^7$ yr 
   on the whole (whatever the timescale of a single lithium enhancement 
   episode and the recurrence of the phenomenon may be); {\it ii)} the 
   photospheric $^7$Li enrichment occurs near the tip of the RGB, so that the 
   mass loss is at work at the highest rates ($\sim$ 10$^{-7}$ $M_\odot$ 
   yr$^{-1}$); {\it iii)} the surface lithium abundance is set to 
   log\,$\epsilon$($^7$Li) = 4 for each star. (Obviously, a reliable 
   computation of the $^7$Li yields from stars in this mass range would be 
   highly desirable.)

\subsubsection{Classical novae}

   In the past decades, contradictory views on lithium production from novae 
   have appeared in the literature. D'Antona \& Matteucci (1991), by using the 
   results of Starrfield et al. (1978) on $^7$Li synthesis in nova outbursts, 
   concluded that novae would represent a significant source for the present 
   ISM $^7$Li content. Later, Boffin et al. (1993) ruled out novae as lithium 
   producers at a Galactic level, but a subsequent check by Hernanz et al. 
   (1996) plotted out again large overproduction factors relative to the solar 
   abundance for $^7$Be (and hence $^7$Li)! 
   More recently, Jos\'e \& Hernanz (1998) computed a grid of hydrodynamical 
   nova models, providing $^7$Li yields which we used in Paper I. In that 
   study, we emphasized the important role played by novae in reproducing the 
   late, steep rise from the lithium-metallicity plateau: since the nova 
   eruptions come from systems containing a cool white dwarf (WD), we have to 
   wait a long time, given by the lifetime of the progenitor star plus a 
   suitable cooling time, before observing any production from these stellar 
   $^7$Li factories. 

   In order to include the nova system nucleosynthesis in the chemical 
   evolution code, we had to make a number of assumptions: {\it i)} the nova 
   system formation rate at any time $t$ is a constant fraction of the WD 
   formation rate at a previous time $t - \Delta\,t$ (D'Antona \& Matteucci 
   1991); this constant fraction is fixed by the rate of nova outbursts 
   observed in the Galaxy at the present time (20\,--\,30 yr$^{-1}$, Shafter 
   1997); {\it ii)} the value of $\Delta\,t$ is set in order to guarantee the
   cooling of the WD at a level that ensures a strong enough nova burst; {\it 
   iii)} 10$^4$ nova outbursts are assumed to occur on average in each nova 
   system (Bath \& Shaviv 1978; Shara et al. 1986); {\it iv)} 30\,\% of novae 
   occur in systems containing ONeMg WDs, while the remaining 70\,\% occur in 
   systems containing CO WDs ({\it e.g.}, Gehrz et al. 1998).

   In this study we will adopt the same nucleosynthesis prescriptions on 
   lithium production from nova outbursts we used in Paper I. Besides, we 
   will investigate in more detail how tightly our results are bound to the 
   assumptions we made.

\subsubsection{GCR nucleosynthesis}

   $^7$Li is also produced in spallation and fusion processes associated with 
   cosmic rays ({\it e.g.}, Reeves et al. 1970; Meneguzzi et al. 1971). 

   A GCR production of $^7$Li at early times could complicate the derivation
   of the primordial $^7$Li abundance from Population II star data also under 
   the simplifying hypothesis that $^7$Li depletion in halo stars is 
   negligible. In principle, by using information on Be and $^6$Li one can set 
   an upper bound to the fraction of $^7$Li produced by GCRs in the halo 
   phase, in the context of a given model of GCR nucleosynthesis (see Olive \& 
   Fields 1999 and refs. therein). However, theoretical predictions are very 
   uncertain, since they depend on the details of the cosmic ray sources and 
   propagation, which are still poorly known. At epochs when [Fe/H] exceeds 
   about $-$\,1 dex, nucleosynthesis in a variety of other Galactic objects 
   (see previous paragraphs) produce the bulk of $^7$Li. In particular, GCRs 
   are expected to contribute no more than 25\,\% of the meteoritic $^7$Li 
   abundance (see arguments in Reeves 1993). 

   In this work we will use the absolute yields from GCRs by Lemoine, 
   Vangioni-Flam, \& Cass\'e (1998) which we already used in Paper I.

\subsubsection{Low-mass X-ray binaries}

   The high Li abundances (20\,--\,200 times the solar value) observed in 
   the low-mass secondaries of several soft X-ray transients (SXTs) 
   (Mart\' \i n et al. 1992, 1994a; Marsh, Robinson, \& Wood 1994; Filippenko, 
   Matheson, \& Barth 1995; Harlaftis, Horne, \& Filippenko 1996; Mart\' \i n 
   et al. 1996) imply a source of recent Li production in these systems 
   (Mart\' \i n et al. 1994a; Mart\' \i n, Spruit, \& van Paradijs 1994b). 

   SXTs, also referred to as X-ray novae, are a subclass of LMXBs which are 
   characterized by strong outbursts lasting several weeks, followed by long 
   quiescent periods ranging from several months to tens of years. There are 
   three main scenarios for the formation of these X-ray transients: {\it i)} 
   initially, the system is a wide binary composed of a very massive primary 
   and a solar-type secondary. When the primary evolves off the main sequence, 
   the secondary is engulfed in the giant's atmosphere. The system enters a 
   common envelope phase which ends with core collapse and SN explosion, or 
   nonexplosive formation of a black hole. {\it ii)} Accretion-induced 
   collapse of a massive WD. {\it iii)} The low-mass secondary is captured by 
   the compact object.

   Mart\' \i n et al. (1994a) have argued that energetic processes associated 
   with the accretion onto the compact object produce the fast particles 
   responsible for Li production. Li would be produced in the accretion disk 
   by spallation of CNO nuclei or by $\alpha$-$\alpha$ reactions and might 
   leave the system through a disk-fed wind; some of the disk wind would then 
   be captured by the secondary, enriching it with Li (Mart\' \i n et al. 
   1994b and refs. therein). The abundances observed in the late-type 
   secondaries one month or longer after a burst must represent closely the 
   average production over many outbursts (Mart\' \i n et al. 1994b). 
   Therefore, from the amount of Li observed at the surface of the secondary 
   one can in principle infer the number of Li atoms created during the 
   outburst that escape and enrich the ISM ({\it e.g.}, Mart\' \i n et al. 
   1994b; Yi \& Narayan 1997).

   To estimate the role these transient systems play as producers of lithium 
   at a Galactic scale we assume that: {\it i)} at any time $t$, the LMXB 
   formation rate is a constant fraction of the Type II SN rate (with Type II 
   SN progenitors in the mass range 8\,--\,100 $M_\odot$). This constant 
   fraction is fixed by the request that the total number of LMXBs that are 
   in quiescence at present is $\sim$ 1000, as estimated by Tanaka \& 
   Shibazaki (1996) for an average recurrence period of 50 yr. However, note 
   that this number could be as low as $\sim$ 200. {\it ii)} The rate at 
   which Li is ejected into the ISM by a single LMXB is of the order of $\sim$
   10$^{\mathbf {-13}}$ $M_\odot$ yr$^{\mathbf {-1}}$, which is the rate of Li 
   production from outbursts for black hole SXTs obtained by using quite 
   optimistic estimates of the various parameters involved in the computation 
   (Yi \& Narayan 1997). However, it should be noted that each neutron star 
   SXT would actually eject Li at a rate about an order of magnitude lower. 
   What's more, we are completely neglecting the fact that during quiescence 
   the rate of Li ejection comes down to lower values. 
   {\it iii)} Each LMXB produces Li for $\sim$ 10 billion years; then, the 
   secondary ends up as a WD and the compact object cannot accrete material 
   any longer. Note that if the mass of the secondary is in the range 
   2\,--\,3 $M_\odot$, Li production should last for shorter periods, ranging 
   from $\sim$ 1.5 Gyr to a few Myrs.

\section{Results}

\subsection{Are AGB stars contributing $^7$Li at a Galactic level?}

   The HBB process has been demonstrated to be able to explain the high $^7$Li 
   abundances observed in the most Li-rich stars which undergo the AGB phase, 
   if coupled to a mixing mechanism (ordinary convective diffusion is shown to 
   work) which takes the newly synthesized $^7$Be to a cooler, external region 
   where it can decay and survive as $^7$Li ({\it e.g.}, Sackmann \& Boothroyd 
   1992). 
   It has also been stated that these Li-rich stars might enrich the ISM with 
   significant amounts of $^7$Li and represent, perhaps, the major source of 
   $^7$Li in a galaxy (Smith \& Lambert 1990; D'Antona \& Matteucci 1991; 
   Sackmann \& Boothroyd 1992; Matteucci et al. 1995). However, detailed 
   computations giving as a result the net $^7$Li yields for a model grid in 
   stellar mass and metallicity (see Section 2.2.2) argue against previous 
   findings and suggest that AGB stars cannot enrich the interstellar matter 
   with significant amounts of lithium, in spite to the fact that large $^7$Li 
   abundances are indeed reached in the external layers at some stages of the 
   evolution (Ventura et al. 2000). To prove this, one needs to include the 
   model grid of $^7$Li yields in a code for the chemical evolution of the 
   solar vicinity, and this is what we have done.

   \begin{figure}
			   \psfig{figure=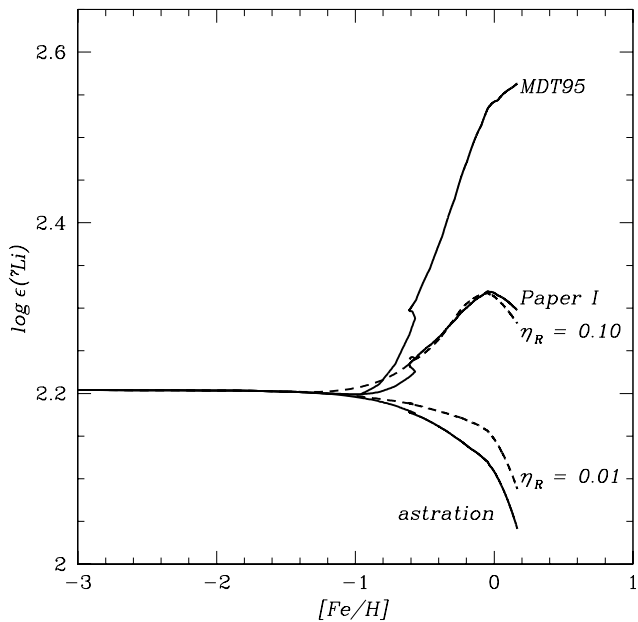,width=8.8cm}
			   \caption{{\it Models labelled MDT95 and Paper I}: 
				    the prescriptions of Matteucci et al. 
				    (1995) on $^7$Li production from AGB stars 
				    are adopted -- $^7$Li production is 
				    allowed only at metallicities $Z$ $>$ 
				    10$^{-3}$; the mean $^7$Li abundance in 
				    the ejecta is log\,$\epsilon$($^7$Li) = 
				    4.15 ({\it MDT95}) or 3.5 ({\it Paper I}).
				    {\it Dashed lines} are models which 
				    include the grid of stellar yields given 
			            in Table~\ref{TabYields}, for values of 
				    the Reimer's factor $\eta_R$ equal to 0.1 
				    or 0.01, respectively. As the lower value 
				    of $\eta_R$ is the preferred one, we can 
			      	    conclude that the contribution from AGB 
				    stars to the overall Galactic $^7$Li 
				    enrichment is negligible (compare results 
				    on AGB stars to the case of pure
				    astration shown on the bottom of the 
				    figure).}
      \label{FigAgb}%
   \end{figure}

   \begin{figure}
			   \psfig{figure=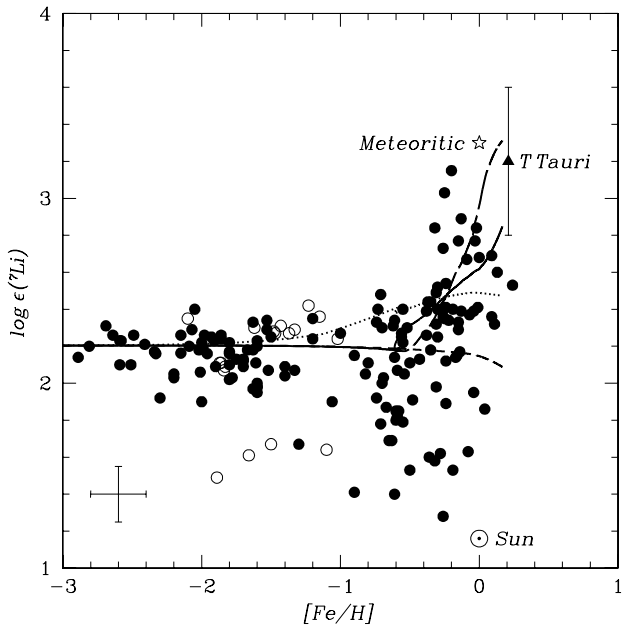,width=8.8cm}
			   \caption{log\,$\epsilon$($^7$Li) vs. [Fe/H] 
				    theoretical trends predicted by different 
				    models compared to observational data of 
				    Fig.1. Each model takes into account 
				    $^7$Li synthesis in a single category of 
				    stellar producers: {\it dots}: Type II 
				    SNe; {\it short-dashed line}: AGB stars 
		 		    (case $\eta_R$ = 0.01); {\it continuous 
				    line}: novae; {\it long-dashed line}: 
				    low-mass giants.}
      \label{FigDif}%
   \end{figure}

   \begin{figure}
			    \psfig{figure=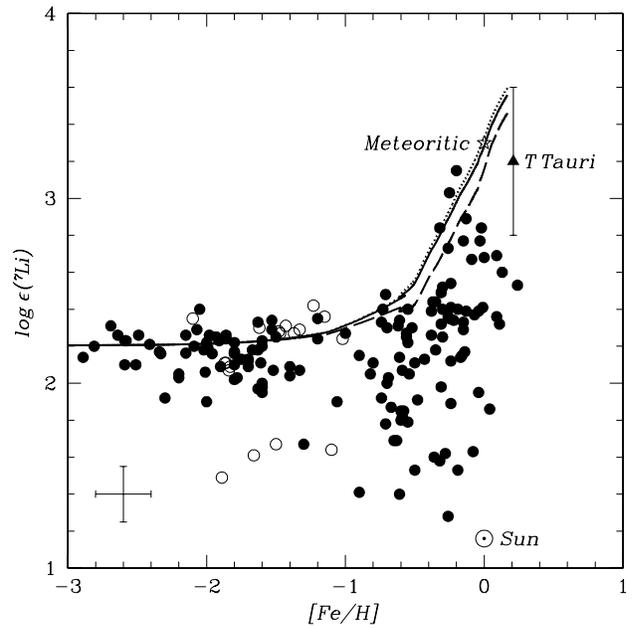,width=8.8cm}
			    \caption{log\,$\epsilon$($^7$Li) vs. [Fe/H] 
				     theoretical trends predicted by models 
				     including all the following astrophysical 
  				     sites for $^7$Li production: {\it dashed 
				     line}: Type II SNe $+$ AGB stars $+$ 
				     low-mass giants $+$ novae; 
				     {\it continuous line}: Type II SNe $+$ 
				     AGB stars $+$ low-mass giants $+$ novae 
				     $+$ GCRs. This latter is our best-model. 
				     Adding the contribution from LMXBs 
				     as discussed in the text does not change 
				     much the result ({\it dotted line}).}
      \label{FigBest}%
   \end{figure}

   In Fig.~\ref{FigAgb} we show the effect of employing the new vs. the old 
   prescriptions on $^7$Li production from AGB stars in models where these 
   stars are the only $^7$Li producers. The curves labelled {\it MDT95} and 
   {\it Paper I} refer to theoretical trends obtained by using the old 
   prescriptions from Matteucci et al. (1995) and Paper I, respectively. In 
   particular, in Paper I we used a conservative estimate of the PN mass 
   ejected by AGB stars:
	\begin{displaymath}
		M_{PN} = 0.8 \times M_{init} - 1.1
	\end{displaymath}
   as in Matteucci et al. (1995). This formula accounts for any mass loss by 
   winds prior to the ejection of the PN. The abundance of $^7$Li in the 
   ejected material was assumed to be log\,$\epsilon$($^7$Li) = 3.5 dex, 
   fairly lower than the value suggested by Matteucci et al. (1995) (4.15 
   dex). In both models $^7$Li production in AGB stars takes place only at 
   metallicities $Z$ $>$ 10$^{-3}$. When the thermally pulsing AGB phase is 
   followed by means of self-consistent stellar evolutionary tracks and mass 
   loss is taken into account in a detailed way, more reliable $^7$Li yields 
   can be obtained. Chemical evolution results produced by using the model 
   grid of Table~\ref{TabYields} for a value of the Reimer's parameter of mass 
   loss $\eta_R$ = 0.10 are very similar to those obtained with our old 
   prescriptions ({\it cf.} models labelled {\it Paper I} and {\it $\eta_R$ = 
   0.10}). On the other hand, the choice $\eta_R$ = 0.01 results in a 
   significantly lower $^7$Li production from AGBs. A value of $\eta_R$ = 0.01 
   seems to be the preferred one in order to explain some important stellar 
   observational properties (Ventura et al. 2000); hence we conclude that 
   $^7$Li production from AGB stars should be negligible and nearly 
   undistinguishable from a situation where only astration is acting and no 
   astrophysical sites for lithium production are turned on (model labelled 
   {\it astration} in Fig.~\ref{FigAgb}).

   In Fig.~\ref{FigDif} we show the $^7$Li abundance evolution that would be 
   predicted by taking into account only a single category of stellar $^7$Li 
   producers (Type II SNe, AGB stars, novae, and low-mass giants), and compare 
   it to the observational data. Lithium synthesis prescriptions for each 
   class of sources are those given in Section 2.2. The sudden increase in 
   log\,$\epsilon$($^7$Li) at [Fe/H] $\sim$ $-$\,0.5 pointed out by the data 
   requires important production at Galactic ages larger than $\sim$ 2.5 Gyr, 
   so that a low-mass stellar component must enter in the model. It is 
   immediately seen how novae and low-mass giants, contributing to $^7$Li 
   enrichment on longer timescales than the other sources, are the best 
   candidates for explaining this late, sudden rise from the 
   lithium-metallicity plateau. 
   As far as the low-mass giants are concerned, the delayed lithium production 
   is essentially due to the long lifetimes of their progenitors whereas, 
   when considering novae, the late lithium enrichment is partly due to the 
   addition of a suitable cooling time to the lifetime of the WD progenitor, 
   which allows the newly formed WD to cool at a level that ensures strong 
   enough nova outbursts. We checked the model behaviour under different 
   assumptions on the value of the cooling time, and found that no relevant 
   changes in the results are produced for values lying in the range 
   $\Delta\,t$ $\sim$ 1\,--\,2 Gyr. Similarly, changing the fraction of the 
   WDs that enter the formation of new nova systems at every time (anyway 
   keeping it always constant over the time), so that the range of values of 
   the present time nova outburst rate as inferred from observations can be 
   reproduced, does not seriously affect the results. In conclusion, as far as 
   the nova contribution is concerned, the major uncertainties reside in the 
   nucleosynthesis computations and in the number of outbursts experienced by 
   the typical nova system.

   Given the low percentage of low-mass Li-rich stars observed in the Galaxy, 
   we did not considered low-mass giants as $^7$Li producers in Paper I. 
   However, both observational and theoretical evidence has grown suggesting 
   that this particular category of stars could eventually represent a 
   non-negligible $^7$Li factory on a Galactic scale (see Section 2.2.3). 
   Therefore, we include now in the code $^7$Li production from low-mass 
   giants as well. We find that a significant contribution from these stellar 
   sources is achieved only under very restrictive hypotheses on the 
   parameters of the lithium enhancement: lithium enhancement has to last for 
   a period of $\sim$ 10$^7$ yr; moreover, the enhancement has to occur in 
   conjunction with strong mass loss rates. This latter requirement can be 
   easily satisfied if lithium enrichment occurs near the tip of the RGB.

   \begin{figure}
			   \psfig{figure=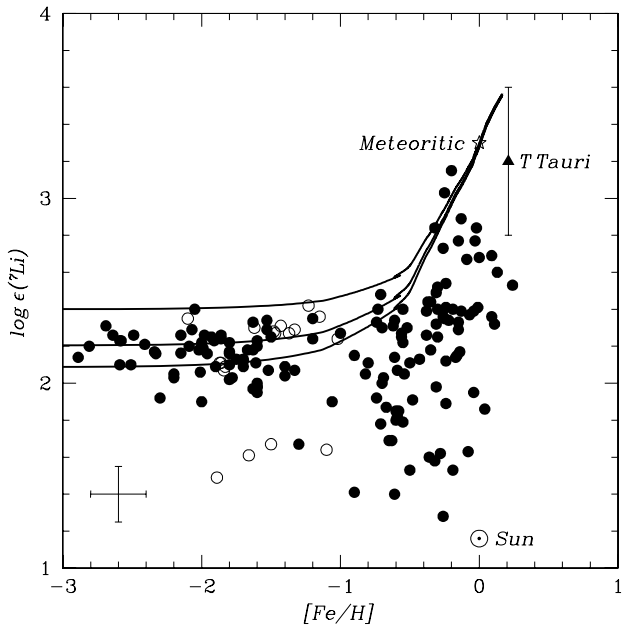,width=8.8cm}
			   \caption{Model predictions are compared to 
			            observational data. We show the effect of 
				    changing the primordial lithium abundance 
				    (from bottom to top: 
				    log\,$\epsilon$($^7$Li)$_P$ = 2.09, 2.2, 
				    and 2.4 dex).}
      \label{FigPrim}%
   \end{figure}

   \begin{table}
      \caption{Percentage of meteoritic $^7$Li contributed from each source
	       in the framework of our best-model.}
         \label{MetAbund}
      \[
         \begin{array}{p{0.5\linewidth}l}
            \hline
            \noalign{\smallskip}
	    Source  &  \% \\
            \noalign{\smallskip}
            \hline
            \noalign{\smallskip}
	    Type II SNe   &   9\% \\
 	    AGB stars   &   0.5\% \\
	    low-mass giants$^{\rm a}$   &   41\% \\
	    novae   &   18\% \\
	    GCRs   &   25\% \\
            \noalign{\smallskip}
            \hline
         \end{array}
      \]
\begin{list}{}{}
\item[$^{\rm a}$] 1 $\le$ $M/M_\odot$ $\le$ 2.
\end{list}
   \end{table}
%
%

\subsection{Our best-model}

   By taking into account all the stellar $^7$Li producers seen above in the 
   chemical evolution model, we are able to explain the steep rise from the 
   lithium-metallicity plateau, but we still fail to achieve the meteoritic 
   abundance of $^7$Li (Fig.~\ref{FigBest}, dashed line). Our best-model 
   includes the contributions from Type II SNe, AGB stars, low-mass giants, 
   novae and GCRs (Fig.~\ref{FigBest}, continuous line). 
   Adding the contribution from LMXBs practically does not change the result 
   (Fig.~\ref{FigBest}, dotted line). 
   This is against LMXBs being considered lithium producers on a Galactic 
   scale, although no firm conclusions can be drawn, given the uncertainties 
   involved when dealing with these systems. However, we would like to stress 
   that we adopted here a series of prescriptions aiming at maximizing the 
   role of LMXBs in enriching the ISM with Li, at least in the framework of 
   our present knowledge of these systems (see Section 2.2.6). 
   In Table~\ref{MetAbund} we list the contribution (in \%) to the meteoritic 
   abundance from each source in the framework of our best-model. 
   The joint contributions from Type II SNe, AGBs, and GCRs are able to 
   explain $\sim$ 35\,\% of the meteoritic $^7$Li abundance. The remainder 
   should be produced by long living sources, which we identify with novae and 
   low-mass giants. However, the relative importance of their contributions is 
   very uncertain. We tentatively ascribe $\sim$ 20\,\% of the meteoritic 
   lithium to nova production and $\sim$ 40\,\% to low-mass giants, on the 
   ground of current knowledge of nova nucleosynthesis and current 
   (un)knowledge of $^7$Li synthesis along the RGB in low-mass stars.

   Finally, we compare our best-model with other two models which share with 
   it the same prescriptions on $^7$Li production, but start from different 
   values of the primordial lithium abundance (Fig.~\ref{FigPrim}). It is 
   worth noticing that model results are nearly the same for metallicities 
   greater than [Fe/H] $\sim$ $-$\,0.5. This is due to the fact that stellar 
   astration acting during the Galaxy evolution strongly reduces the 
   primordial $^7$Li abundance in the ISM already after $\sim$ 2.5 Gyr from 
   the beginning of the Galaxy formation\footnote{In the framework of the 
   model presented here, an ISM metallicity of [Fe/H] $\sim$ $-$\,0.5 is 
   gained at $t$ $\sim$ 2.5 Gyr.}. For $t$ $>$ 2.5 Gyr only a small fraction 
   of the observed $^7$Li has a primordial origin, since at that time most of 
   the ISM $^7$Li abundance is due to stellar production.

\section{Conclusions}

\begin{itemize}
\item In agreement with our previous results (Paper I) we confirm 
      that novae are necessary in order to reproduce the 
      $^7$Li abundance evolution traced out by the upper envelope of the 
      observational data; in particular, they can reproduce the steep rise 
      of the lithium abundance for [Fe/H] $\ge$ $-$\,0.5 dex.
\item New stellar yields of $^7$Li from AGB stars have been included in the 
      code for the chemical evolution of the Galaxy. As a result, we find that 
      AGB stars cannot be considered as a significant source of $^7$Li in the 
      Galaxy.
\item Low-mass giant stars, restoring their processed material on long 
      timescales, are among the best candidates (together with novae) for 
      reproducing the late rise from the lithium-metallicity plateau. Recent 
      theoretical computations (Sackmann \& Boothroyd 1999) suggest that they 
      could indeed contribute a non-negligible fraction of the interstellar 
      $^7$Li, and this is confirmed by our model results. However, it should 
      be cautioned that in order to reproduce the observed 
      log\,$\epsilon$($^7$Li) vs. [Fe/H] trend we have to make very 
      restrictive hypotheses on the parameters of the lithium enhancement in 
      low-mass RGB stars.
\item As far as the LMXBs are concerned, we conclude that they contribute 
      negligibly to the amount of lithium in the Galaxy.
\item The temporal evolution of the $^7$Li abundance in the disk is almost 
      independent of the value of the primordial abundance we choose. On 
      the contrary, the evolutionary scenario in the halo is deeply bound to 
      the choice of this value.
\item Our best-model predicts an increase in the ISM $^7$Li content over the 
      last 4.5 Gyr. At first glance, this seems to be in disagreement with 
      recent observations both of T\,Tauri stars (Stout-Batalha et al. 2000) 
      and in the Orion association (Cunha, Smith, \& Lambert 1995). However, 
      as far as the Orion stars are concerned, there are arguments for 
      accommodating a $^7$Li abundance near-solar even if the Galactic disk 
      $^7$Li abundances are increasing (Cunha et al. 1995): in particular, the 
      fact that the mean [Li/Fe] for the Orion stars is roughly 0.0 to 
      $-$\,0.1, and therefore not very far from the meteoritic value, would be 
      indicative of the fact that the $^7$Li content measured in the Orion 
      association should not be compared to our results for the $^7$Li 
      abundance in the ISM at the present time, but with those referring to 
      4.5 Gyr ago.
\end{itemize}
   Finally, we would like to recall the reader that all our computations have 
   been carried out under the assumption that the stars tracing the upper 
   envelope of the log\,$\epsilon$($^7$Li) vs. [Fe/H] diagram have not had 
   their original lithium abundance altered by any process of dilution and/or 
   depletion.

\begin{acknowledgements}
We acknowledge an anonymous referee for the helpful criticism.
\end{acknowledgements}

\end{document}